\documentclass[pra,twoside,twocolumn,showpacs,superscriptaddress,floatfix]{revtex4}
\usepackage{times,graphicx,bbm,amsmath,amssymb}
\usepackage{epsfig,color}
\newcommand{\bmsigma}{\boldsymbol \sigma} 
\newcommand{\bmLambda}{\boldsymbol \Lambda} \def\D{{\rm D}}
\newcommand{\bmlambda}{\boldsymbol \lambda} \def\X{\boldsymbol{X}}
\def\Tr{\hbox{Tr}} \def\sigmaCM{\boldsymbol{\sigma}}
\begin{document}
\title{A measure of the non-Gaussian character of a quantum state}
\author{Marco G. Genoni}
\affiliation{Dipartimento di Fisica dell'Universit\`a di Milano,
I-20133, Milano, Italia.}
\author{Matteo G. A. Paris}
\email{matteo.paris@fisica.unimi.it}
\affiliation{Dipartimento di Fisica dell'Universit\`a di Milano,
I-20133, Milano, Italia.}
\affiliation{Institute for Scientific Interchange, I-10133 Torino, Italia}
\author{Konrad Banaszek}
\affiliation{Institute of Physics, Nicolaus Copernicus University,
PL-87-100 Toru\'{n}, Poland}
\date{\today}
\begin{abstract}
We address the issue of quantifying the non-Gaussian character of
a bosonic quantum state and introduce a non-Gaussianity measure
based on the Hilbert-Schmidt distance between the state under
examination and a reference Gaussian state. We analyze in details
the properties of the proposed measure and exploit it to evaluate
the non-Gaussianity of some relevant single- and multi-mode
quantum states.  The evolution of non-Gaussianity is also
analyzed for quantum states undergoing the processes of
Gaussification by loss and de-Gaussification by
photon-subtraction.  The suggested measure is easily computable
for any state of a bosonic system and allows to define a
corresponding measure for the non-Gaussian character of a quantum
operation.
\end{abstract}
\maketitle
\section{Introduction}\label{s:intro}
Gaussian states play a crucial role in quantum information
processing with continuous variables. This is especially true
for quantum optical implementations since radiation at thermal
equilibrium, including the vacuum state, is itself a Gaussian
state and most of the Hamiltonians achievable within the current
technology are at most bilinear in the field operators, {\em i.e.}
preserve the Gaussian character \cite{AOP,Eisert,Illu}.
As a matter of fact, using single-mode and entangled Gaussian states,
linear optical circuits and Gaussian operations, like homodyne
detection, several quantum information protocols have been
implemented, including teleportation, dense coding and quantum
cloning \cite{Brau}.
\par
On the other hand quantum information protocols required for
long distance communication, as for example entanglement
distillation and entanglement swapping,  rely on non-Gaussian
operations.  In addition, it has been demonstrated that teleportation
\cite{Tom,IPS2a,IPS2b} and cloning \cite{nonGclon} of quantum
states may be improved by using non-Gaussian states and non-Gaussian
operations. Indeed, de-Gaussification protocols for
single-mode and two-mode states have been proposed
\cite{Tom,IPS2a,IPS2b} and realized \cite{IPS_Wenger}.
It should be also noticed that any strongly
superadditive function is minimized, at fixed covariance
matrix, by Gaussian states. This is crucial to prove
extremality of Gaussian states and Gaussian
operations \cite{Wolf1,Wolf2} for what concerns various
quantities as channel capacities \cite{HW01}, multipartite
entanglement measures \cite{EM} and distillable secret key in
quantum key distribution protocols.
Since in most cases these quantities can be computed only for
Gaussian states, a non-Gaussianity measure may serve as a
guideline to quantify them for the class of non-Gaussian states.
Overall, non-Gaussianity is revealing itself as a resource for continuous
variable quantum information, and thus we urge a measure able to
quantify the non-Gaussian character of a quantum state.
\par
In this paper we introduce a novel quantity, the non-Gaussianity
$\delta [\varrho]$ of a quantum state, which quantifies how much
a state fails to be Gaussian. Our measure, which is based on the
Hilbert-Schmidt distance between the state itself and a reference
Gaussian state, can be easily computed for any state, either
single-mode or multi-mode.
\par The paper is structured as follows. In the next Section
we introduce notation and review the basic properties of Gaussian
states. Then, in Section \ref{s:nong} we introduce
the formal definition of $\delta[\varrho ]$ and study its
properties in details. In Section \ref{s:zoo} we evaluate non-Gaussianity
of relevant quantum states whereas in Section \ref{s:pro} we analyze the
evolution of non-Gaussianity for known Gaussification and
de-Gaussification maps. Section \ref{s:out} closes the paper with some
concluding remarks.
\section{Gaussian states}
For concreteness, we will use here the quantum optical terminology
of modes carrying photons, but our theory applies to general bosonic
systems.
Let us consider a system of $n$ modes described by mode
operators $a_k$, $k=1\dots n$, satisfying the commutation relations
$[a_k,a_j^{\dag}]=\delta_{kj}$. A quantum state $\varrho$ of the
$n$ modes is fully described by its characteristic function
\cite{Glauber2}
$$
\chi[\varrho](\bmlambda) = \Tr[\varrho\,D(\bmlambda)]
$$
where $D(\bmlambda) = \bigotimes_{k=1}^n D_k(\lambda_k)$
is the $n$-mode displacement operator, with $\bmlambda =
(\lambda_1,\dots,\lambda_n)^T$, $\lambda_k \in \mathbbm{C}$, and where
$$
D_k(\lambda_k) =\exp\{\lambda_k a_k^{\dag} - \lambda_k^* a_k \}
$$
is the single-mode displacement operator.
The canonical operators are given by:
\begin{align}
q_k &= \frac{1}{\sqrt{2}}(a_k + a^{\dag}_k), \nonumber \\
p_k &= \frac{1}{i\sqrt{2}}(a_k - a_k^{\dag}) \nonumber
\end{align}
with commutation relations given by $[q_j,p_k]=i\delta_{jk}$.
Upon introducing the  real vector $\boldsymbol{R}=(q_1,p_1,\dots,q_n,p_n)^T$,
the commutation relations rewrite as
$$
[R_k,R_j] = i \Omega_{kj}
$$
where $\Omega_{kj}$ are the elements of the symplectic matrix
$\boldsymbol{\Omega} = i \bigoplus_{k=1}^n \sigma_2$, $\sigma_2$
being the $y$-Pauli matrix.
The covariance matrix $\bmsigma\equiv\bmsigma[\varrho]$ and the vector
of mean values $\X \equiv \X[\varrho]$ of a quantum state
$\varrho$ are defined as
\begin{align}
{X}_j &= \langle R_j \rangle \nonumber \\
\sigma_{kj} &= \frac{1}{2}\langle \{ R_k,R_j \}\rangle - \langle R_j
\rangle\langle R_k\rangle \nonumber \\
\end{align}
where $\{ A,B \} = AB + BA$ denotes the anti-commutator, and
$\langle O \rangle = \Tr[\varrho\:O]$ is the expectation value
of the operator $O$.
\par
A quantum state $\varrho_G$ is referred to as a Gaussian state if
its characteristic function has the Gaussian form
$$
\chi[\varrho_G](\bmLambda) = \exp \left\{ - \frac{1}{2}
\bmLambda^T \bmsigma \bmLambda + \X^T \boldsymbol{\Omega} \bmLambda \right\}
$$
where $\bmLambda$ is the real vector $\bmLambda = (\hbox{Re}
\lambda_1, \hbox{Im}\lambda_1, \dots, \hbox{Re} \lambda_n,
\hbox{Im}\lambda_n)^T$. Of course, once the covariance matrix and
the vector of mean values are given, a Gaussian state is fully
determined. For a single-mode system the most general
Gaussian state can be written as
$$
\varrho_G= D(\alpha) S(\zeta)
\nu(n_t) S^\dag (\zeta) D^\dag(\alpha) ,
$$
$D(\alpha)$ being the
displacement operator, $S(\zeta) = \exp[\frac12 \zeta (a^{\dag})^2
- \frac12 \zeta^* a^2]$ the squeezing operator, $\alpha,\zeta \in
{\mathbbm C}$, and $\nu(n_t)=(1+n_t)^{-1} [n_t/(1+n_t)]^{a^\dag
a}$ a thermal state with $n_t$ average number of photons.
\section{A measure of the non-Gaussian character of a quantum state}
\label{s:nong}
In order to quantify the non-Gaussian character of a quantum
state $\varrho$ we use a quantity based on the distance between
$\varrho$ and a reference Gaussian state $\tau$,
which itself depends on $\varrho$.
 Specifically, we define the non-Gaussianity $\delta[\varrho]$ of
the state $\varrho$ as
\begin{align}
\delta[\varrho] = \frac{ \D_{HS}^2[\varrho,\tau] } {\mu[\varrho]} \label{eq:nonG}
\end{align}
where $\D_{HS}[\varrho,\tau]$ denotes the Hilbert-Schmidt distance between
$\varrho$ and $\tau$
\begin{align}
\D_{HS}^2[\varrho,\tau] &= \frac{1}{2} \Tr[(\varrho - \tau)^2] =
\frac{ \mu[\varrho] + \mu[\tau] - 2\kappa[\varrho,\tau] }{2},
\end{align}
with $\mu[\varrho]=\hbox{Tr}[\varrho^2]$ and
$\kappa[\varrho,\tau]=\hbox{Tr}[\varrho\:\tau]$
denoting the purity of $\varrho$ and the overlap between
$\varrho$ and $\tau$ respectively. The Gaussian reference $\tau$
is the Gaussian state such that
\begin{align}
\X[\varrho] &= \X[\tau] \nonumber \\
\boldsymbol{\sigma}[\varrho] &= \boldsymbol{\sigma}[\tau] \nonumber
\end{align}
\emph{i.e.} $\tau$ is the Gaussian state with the same covariance matrix
$\bmsigma$ and the same vector $\X$ of the state $\varrho$.
\par
The relevant properties of $\delta [\varrho ]$, which confirm
that it represents a good measure of the non-Gaussian character
of $\varrho$, are summarized by the following Lemmas:
\par\noindent
{\bf Lemma 1}:  $\delta[\varrho]=0$ iff $\varrho$ is a Gaussian state.
\par\noindent
{\bf Proof}: If $\delta[\varrho]=0$ then $\varrho=\tau$ and thus it is
a Gaussian state. If $\varrho$ is a Gaussian state, then it is uniquely
identified by its first and second moments and thus the reference Gaussian
state $\tau$ is given by $\tau=\varrho$, which, in turn, leads to
$\D_{HS}[\varrho,\tau]=0$ and thus to $\delta[\varrho]=0$.
\par\noindent
{\bf Lemma 2}: If $U$ is a unitary map corresponding to a symplectic
transformation in the phase space, \emph{i.e.} if $U=\exp\{-i H\}$ with
 hermitian $H$ that is at most bilinear in the field operators, then
$\delta[U\varrho U^{\dag}] = \delta[\varrho]$.
This property ensures that displacement and squeezing operations do
not change the Gaussian character of a quantum state.
\par\noindent
{\bf Proof}:  Let us consider $\varrho^\prime = U\varrho U^\dag$.
Then the
covariance matrix transforms as $\sigmaCM[\varrho^\prime] =
\Sigma \sigmaCM[\varrho] \Sigma^T$, $\Sigma$ being the symplectic
transformation associated to $U$. At the same time the vector of
mean values simply translates to $\X^{\prime}=\X+\X_0$, where $\X_0$
is the displacement generated by $U$. Since any
Gaussian state is fully characterized by its first and second
moments, then the reference state must necessarily transform as
$\tau^\prime = U \tau U^\dag$, \emph{i.e.} with the same unitary
transformation $U$. Since the Hilbert-Schmidt distance and the purity
of a quantum state are invariant under unitary transformations
the lemma is proved.
\par\noindent
{\bf Lemma 3}: $\delta[\varrho]$ is proportional to the squared
$L^2(\mathbbm{C}^n)$ distance between the characteristic functions of
$\varrho$ and of the reference Gaussian state $\tau$.
In formula:
\begin{align}
\delta[\varrho] \propto \int d^{2n}{\bmlambda} \: [
\chi[\varrho](\bmlambda) - \chi[\tau](\bmlambda)]^2 \:.
\end{align}
Since the notion of Gaussianity of a quantum state is defined
through the shape of its characteristic function, and since the
characteristic function of a quantum state belongs to the
$L^2(\mathbbm{C}^n)$ space \cite{Glauber2}, we address
$L^2(\mathbbm{C})$ distance to as a good indicator for the non
Gaussian character of $\varrho$.
\par\noindent
{\bf Proof}:  Since characteristic functions of self-adjoint operators are
even functions of $\lambda$  and by means of the identity
$$\Tr[O_1O_2] = \int \frac{d^{2n}{\bmlambda}}{\pi^n}
\chi[O_1](\bmlambda)\,\chi[O_2](-\bmlambda)\,,$$
we obtain
$${\D}_{HS}^2[\varrho,\tau] = \frac{1}{2} \int \frac{d^{2n}{\bmlambda}}{\pi^n}\,
\left[ \chi[\varrho](\bmlambda) - \chi[\tau](\bmlambda) \right]^2\:.$$
\par\noindent
{\bf Lemma 4}: Consider a bipartite state $\varrho=\varrho_A\otimes\varrho_G$.
If $\varrho_G$ is a Gaussian state then $\delta[\varrho] = \delta[\varrho_A]$.
\par\noindent
{\bf Proof}: we have
\begin{align}
\mu[\varrho]&= \mu[\varrho_A] \mu[\varrho_G] \nonumber \\
\mu[\tau]&= \mu[\tau_A] \mu[\tau_G] \nonumber \\
\kappa[\varrho,\tau]&=\kappa[\varrho_A,\tau_A]\kappa[\varrho_G,\varrho_G]
\:. \nonumber
\end{align}
Therefore, since $\kappa[\varrho_G,\varrho_G]=\mu[\varrho_G]$ we arrive at
\begin{align}
\delta[\varrho] &=
\frac{\mu[\varrho_A] \mu[\varrho_G] + \mu[\tau_A]\mu[\varrho_G]
- 2 \kappa[\varrho_A,\tau_A]\kappa[\varrho_G,\varrho_G]}{2 \mu[\varrho_A]
\mu[\varrho_G]} \nonumber \\
&= \delta[\varrho_A]
\end{align}
\par
The four properties illustrated by the above lemmas are the
natural properties required for a good measure of the
non-Gaussian character of a quantum state.  Notice that by using
the trace distance $\D_{T}[\varrho,\tau] = \frac{1}{2}
\Tr|\varrho-\tau|$ instead of the Hilbert-Schmidt distance we
would lose Lemmas 3 and 4, and that the invariance expressed by
Lemma 4 holds thanks to the \emph{renormalization} of the
Hilbert-Schmidt distance through the purity $\mu[\varrho]$.  We
stress the fact that our measure of non-Gaussianity is a
computable one: It may be evaluated for any quantum state of $n$
modes by the calculation of the first two moments of the state,
followed by the evaluation of the overlap with the corresponding
Gaussian state.
\par
Notice that $\delta [\varrho]$ is not additive (nor
multiplicative) with respect to the tensor product. If  we
consider a (separable) multi-partite quantum state  in the product form
$\varrho = \otimes_{k=1}^n \varrho_k$,  the non-Gaussianity is
given by
\begin{equation}
\delta[\varrho] = \frac{ \prod_{k=1}^n \mu[\varrho_k] +
\prod_{k=1}^{n}\mu[\tau_k]  - 2 \prod_{k=1}^n \kappa[\varrho_k,\tau_k]} {2
\prod_{k=1}^n \mu[\varrho_k] } \label{eq:nonGMultiPartite}
\end{equation}
where $\tau_k$ is the Gaussian state with the same moments of
$\varrho_k$.  In fact, since the state $\varrho$ is factorisable,
we have that the corresponding Gaussian $\tau$  is a factorisable
state too.
\section{Non-Gaussianity of relevant quantum states}
\label{s:zoo}
\begin{figure}[h!]
\begin{tabular}{r}
\includegraphics[width=0.4\textwidth]{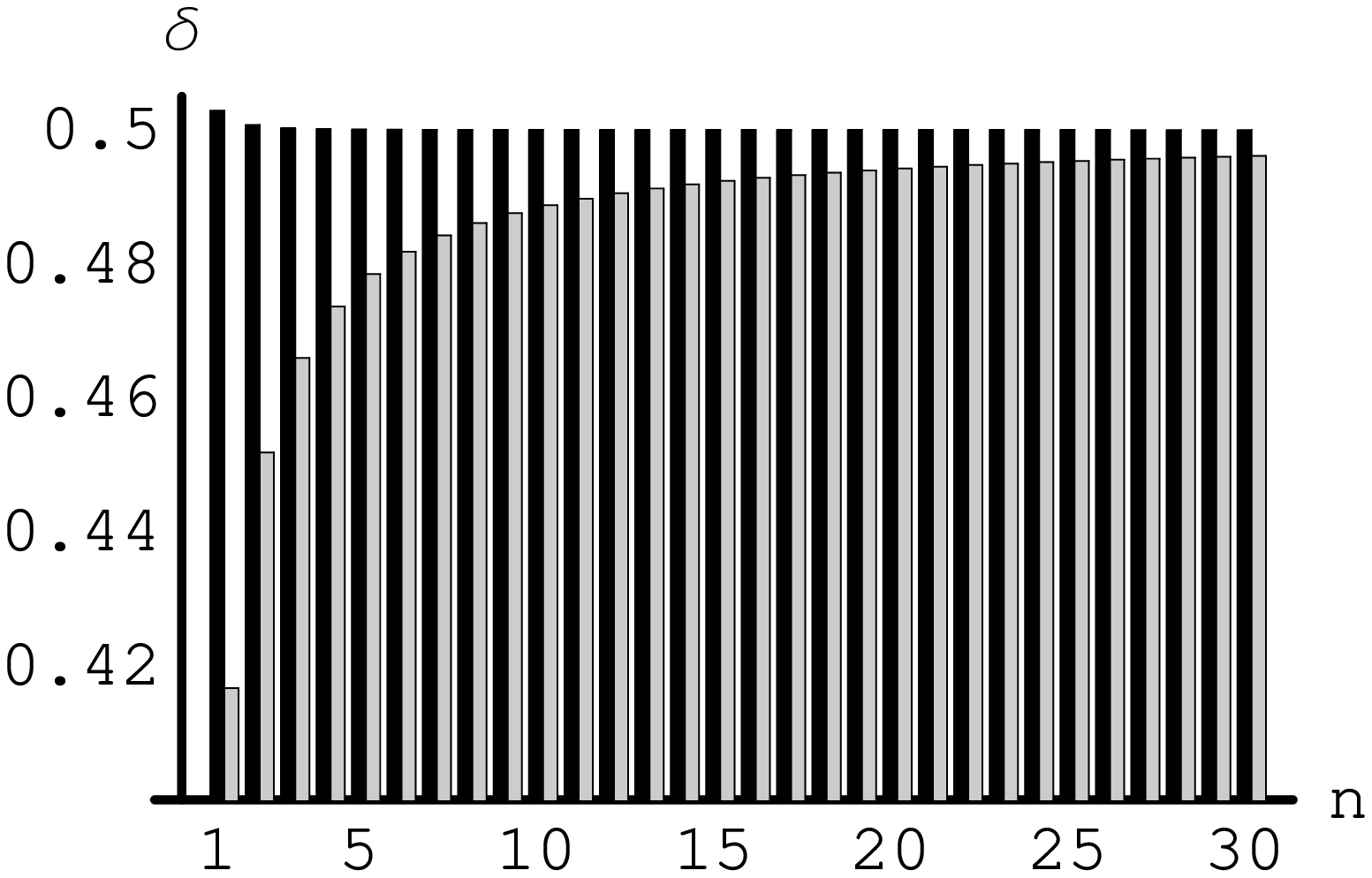} \\
\includegraphics[width=0.4\textwidth]{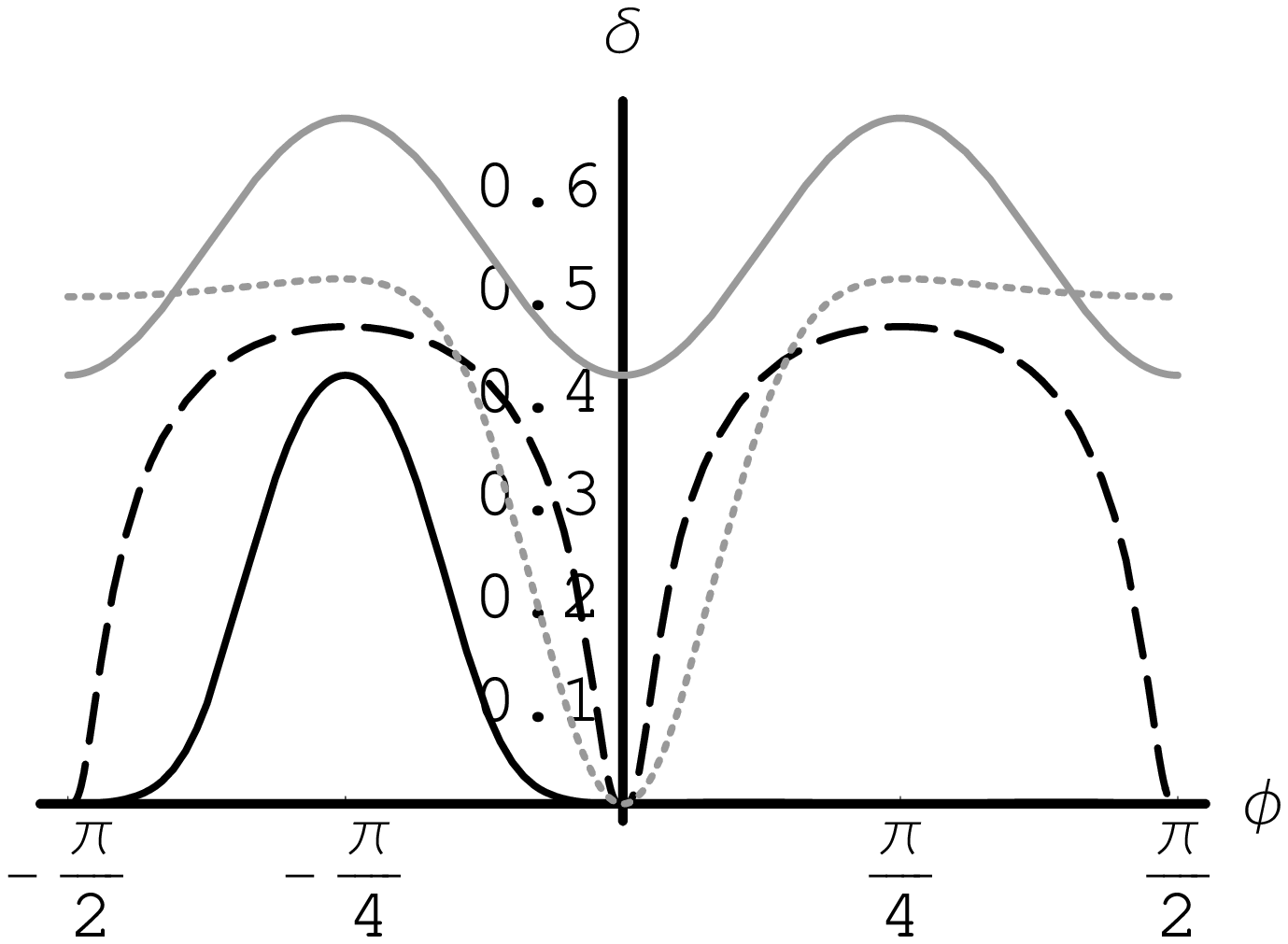}
\end{tabular}
\caption{
\label{f:states}
(Top): Non-Gaussianity of single mode Fock states (gray)
$|p\rangle$ and of multi-mode Fock states $|p\rangle^{\otimes
n}$ (black) as a function of $p$. Non-Gaussianity for multi-mode states
has been maximized over the number of copies $n$. (Bottom):
Non-Gaussianity, as a function of the parameter $\phi$, for the
two-mode superpositions $|\Phi\rangle\rangle$ (dashed gray),
$|\Psi\rangle\rangle$ (solid gray), and for the single-mode
superposition of coherent states $|\psi_S\rangle$ for
$\alpha=0.5$ (solid black) and $\alpha=5$ (dashed black).}
\end{figure}

Let us now exploit the definition (\ref{eq:nonG}) to evaluate the
non-Gaussianity of some relevant quantum states. At first we
consider Fock number states $|p\rangle$ of a single mode as well
as multimode factorisable states  $|p\rangle^{\otimes n}$
made of $n$ copies of a number state.
The reference Gaussian states are a
thermal state $\tau_p = \nu (p)$ with average photon number $p$
and a factorisable thermal state  $\tau_N =  [\nu (p)]^{\otimes n}$
with average photon number $p$ in each mode \cite{mar}.
Non-Gaussianity may be analytically evaluated, leading to
\begin{align}
\delta[|p\rangle\langle p|] &= \frac{1}{2}\left( 1 + \frac{1}{2p +1}
\right) - \frac{1}{p+1} \left(\frac{p}{p+1}\right)^p \nonumber \\
\delta[(|p\rangle\langle p|)^{\otimes n} ] &= \frac{1}{2} \left[ 1 +
\left(\frac{1}{2p +1} \right)^n \right] - \left[ \frac{1}{p+1}\left(
\frac{p}{p+1} \right)^p \right]^n \nonumber
\end{align}
In the multimode
case  of $|p\rangle^{\otimes n}$, we seek for the number of copies that
maximizes the non-Gaussianity. In Fig. \ref{f:states} we
show both $\delta_p\equiv \delta [|p\rangle\langle p|]$ and
$\bar\delta_p= \max_n\delta [(|p\rangle\langle p|)^{\otimes n}]$ as a
function of $p$. As it is apparent from the plot non-Gaussianity
of Fock states $|p\rangle$ increases monotonically with the
number of photon $p$ with the limiting value $\delta_p=1/2$
obtained for $p\rightarrow\infty$. Upon considering multi-mode
copies of Fock states we obtain larger value of non-Gaussianity:
$\bar\delta_p$ is a decreasing function of $p$, approaching
$\bar\delta =1/2$ from above. The value of $\bar\delta_p$
corresponds to $n=3$ for $p < 26$ and to
$n=2$ for $27 \leq p \lesssim 250$.
\par
Another example is the superposition of coherent states
\begin{align}
|\psi_S\rangle &= {\cal N}^{-1/2} \left(
\cos\phi |\alpha\rangle + \sin\phi|-\alpha\rangle
\right) \label{eq:cat}
\end{align}
with  normalization
${\cal N} = 1 + \sin (2\phi) \exp\{-2\alpha^2\}$
which for $\phi=\pm \pi/4$ reduces to the so-called Schr\"odinger
cat states, and whose reference Gaussian state is a displaced
squeezed thermal state $\tau_S = D(C)S(r)\nu (N)
S^{\dag}(r)D^{\dag}(C)$, where the real parameters $C$, $r$, and $N$ are
analytical functions of $\phi$ and $\alpha$. Finally we evaluate
the non-Gaussianity of the two-mode Bell-like superpositions
of Fock states
\begin{align}
|\Phi\rangle\rangle &= \cos\phi |0,0\rangle + \sin\phi
|1,1\rangle \nonumber \\
|\Psi\rangle\rangle &= \cos\phi |0,1\rangle +
\sin\phi |1,0\rangle, \nonumber
\end{align}
which for $\phi=\pm \pi/4$ reduces to the
Bell states $|\Phi^{\pm}\rangle$ and $|\Psi^{\pm}\rangle$. The
corresponding reference Gaussian states are respectively
a two mode squeezed thermal state
$\tau_{\Phi} = S_2(\xi)[\nu (N) \otimes \nu (N)]
S_2^{\dag}(\xi)$, where $S_2(\xi) = \exp (\xi a^\dag_1 a^\dag_2 -
\xi^* a b)$ denotes the two-mode squeezing operator,  and
$\tau_{\Psi} = R(\theta)[ \nu (N_1) \otimes\nu (N_2) ]
R^{\dag}(\theta)$, namely the correlated two-mode state obtained
by mixing a single-mode thermal state with the vacuum at a beam
splitter of transmissivity $\cos^2\theta$, {\em i.e.} $R(\theta)=
\exp[i \theta (a^\dag_1 a_2 + a^\dag_2 a_1)]$. All the
parameters involved in these reference Gaussian states are
analytical functions of the superposition parameter $\phi$.
Non-Gaussianities are thus evaluated by means of (\ref{eq:nonG})
and are reported in Fig. \ref{f:states} as a function of
the parameter $\phi$.
As it is apparent from the plot, the non-Gaussianity of
single-mode states does not surpass the value $\delta=1/2$, and
this fact is confirmed by other examples not reported here.
\par
As concern the cat-like states, we
notice that for small values of $\alpha$ the non-Gaussianity of
the superposition $|\psi_S\rangle$ shows a different behavior for
positive and negative values of the parameter $\phi$: for $\phi
>0$ and $\alpha=0.5$ we have almost zero $\delta$, while higher
values are achieved for $\phi<0$.  For higher values of $\alpha$
($\alpha=5$ in Fig. \ref{f:states}), non-Gaussianity becomes an
even function of $\phi$. This different behavior can be
understood  by looking at the Wigner functions of even and odd
Schr\"odinger cat states for different values of $\alpha$: for
small values of $\alpha$ the even cat's Wigner function is
similar to a Gaussian function, while the odd cat's Wigner
function shows a non-Gaussian hole in the origin of the phase
space; increasing the value of $\alpha$ the Wigner functions of
the two kind of states become similar and deviate from a Gaussian
function.
\par
\begin{figure}[h]
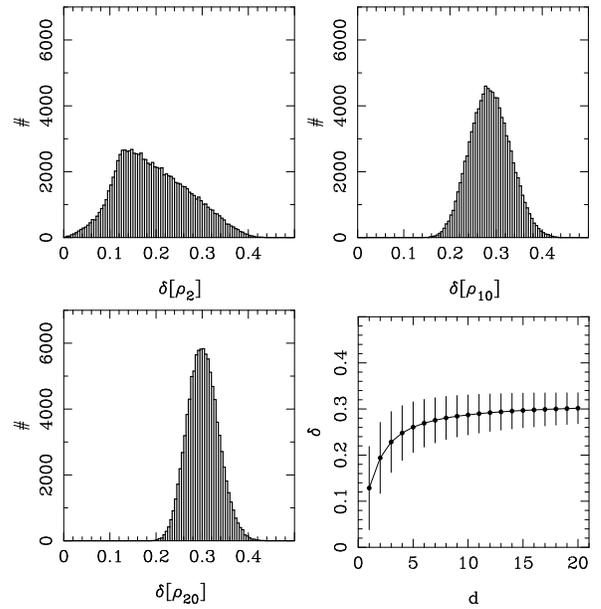

\begin{tabular}{cc}
\includegraphics[width=0.21\textwidth]{f2_a.ps} &
\includegraphics[width=0.21\textwidth]{f2_b.ps} \\
\includegraphics[width=0.21\textwidth]{f2_c.ps} &
\includegraphics[width=0.21\textwidth]{f2_d.ps}
\end{tabular}
\caption{Distribution of non-Gaussianity $\delta[\varrho_d]$
as evaluated for $10^5$ random quantum states, for three different
value of the maximum number of photons $d$. Top: $d=2$ (left), $d=10$ (right);
Bottom: $d=20$ (left). (Bottom-right): Mean values and variances of the
non-Gaussianities evaluated for $10^5$ random quantum states, as a function of the
maximum number of photons $d$. \label{f:random}}
\end{figure}

We have also done a numerical analysis of non-Gaussianity of
single-mode quantum states represented by finite superposition
of Fock states
\begin{align}
\varrho_d = \sum_{n,k=0}^d \varrho_{nk} |n\rangle\langle k |\:.
\end{align}
To this aim we generate randomly quantum states in a finite dimensional
subspaces, $\hbox{dim}(H) \equiv d+1 \leq 21$, following the algorithm
proposed by Zyczkowski {\em et al} \cite{Zyczk1,Zyczk2}, {\em i.e.} by
generating  a random diagonal state ({\em i.e.} a point on the simplex) and
a random unitary matrix according to the Haar measure. In Fig. \ref{f:random}
we report the distribution of non-Gaussianity
$\delta[\varrho_d]$, as evaluated for $10^5$ random quantum states, for
three different value of the
maximum number of photons $d$. As it is apparent from the plots the
distribution of $\delta[\varrho_d]$ becomes Gaussian-like for increasing $d$.
In the fourth panel
of Fig. \ref{f:random} we thus report the mean values and variances of the
the distributions
as a function of the maximum  number of photons $d$.
The mean value increases with the dimension whereas the variance is a
monotonically decreasing function of $d$.
\par
Also for finite superpositions
simulations we did not observe non-Gaussianity higher than $1/2$.
Therefore, although we have no proof, we conjecture that
$\delta=1/2$ is a limiting value for the non-Gaussianity of a
single-mode state. Higher values are achievable for two-mode or
multi-mode quantum states (e.g. $\delta=2/3$ for the Bell states
$|\Psi^{\pm}\rangle\rangle$).

\section{Gaussification and de-Gaussification processes}
\label{s:pro}
\begin{figure}[h!]
\includegraphics[width=0.24\textwidth]{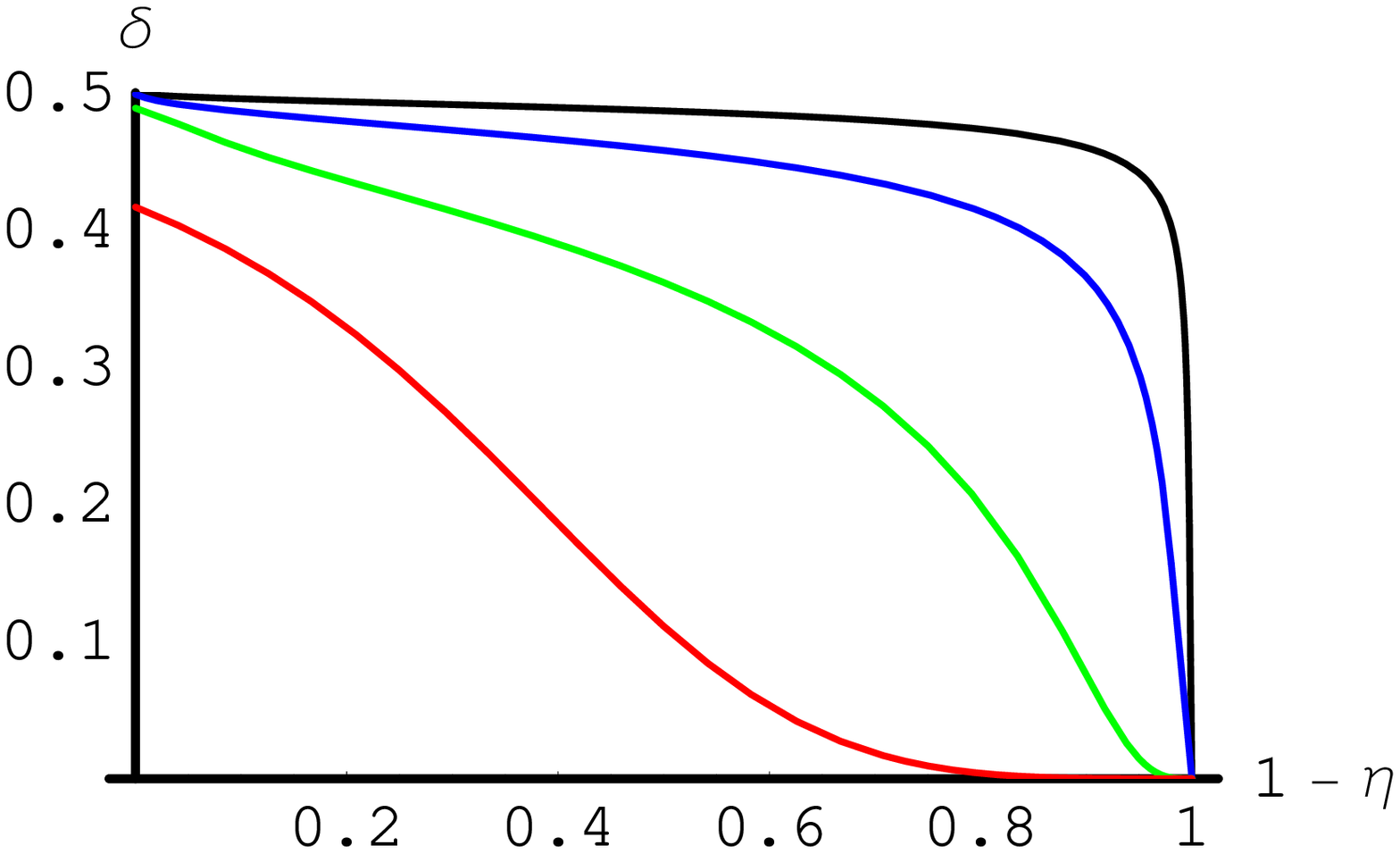}
\includegraphics[width=0.21\textwidth]{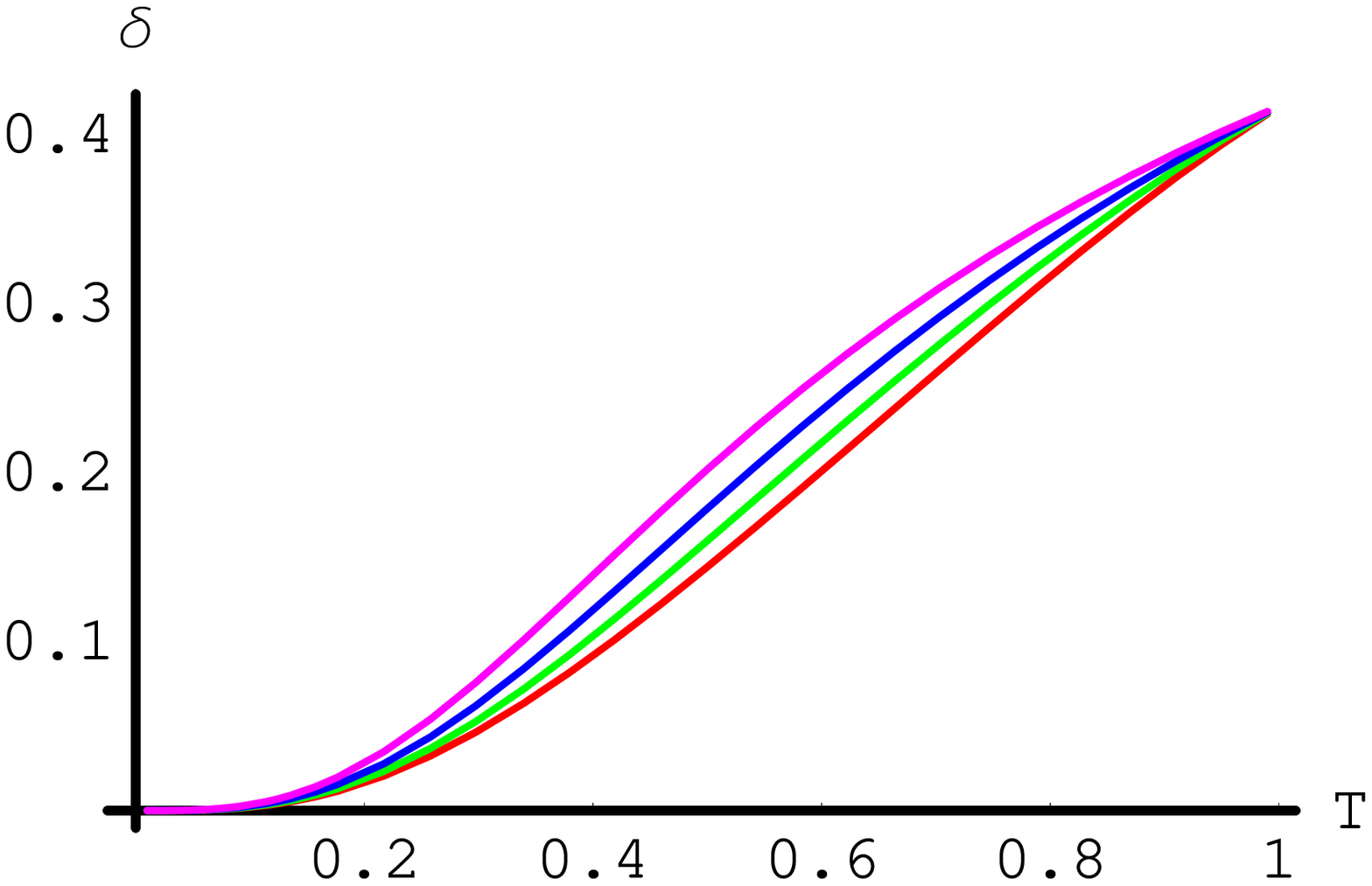}
\caption{
\label{f:procs}
(Left):
Non-Gaussianity of Fock states $|p\rangle$
undergoing Gaussification by loss mechanism due to the
interaction with a bath of oscillators at zero temperature. We
show $\delta_{\eta p}$ as a function of $1-\eta$ for different
values of $p$: from bottom to top $p=1, 10, 100, 1000$.
(Right):
Non-Gaussianity of $\varrho_{IPS}$ as a function of $T$ for
$r=0.5$ and for different values of $\epsilon=0.2, 0.4, 0.6, 0.8$
(from bottom to top). $\delta_{IPS}$ results to be a monotonous
increasing function of $T$, while $\epsilon$ only slightly
changes the non-Gaussian character of the state.
}
\end{figure}

We have also studied the evolution of non-Gaussianity of quantum
states undergoing either Gaussification or de-Gaussification
processes. First we have considered the Gaussification of Fock
states due do the interaction of the system with a bath of
oscillators at zero temperature. This is perhaps the simplest
example of a Gaussification protocol. In fact the interaction
drives asymptotically any quantum state to the vacuum state of
the harmonic system, which, in turn, is a Gaussian state.  The
evolution of the system is governed by the Lindblad Master
equation $\dot{\varrho}=\frac{\gamma}{2}\mathcal L[a]\varrho$,
where $\dot{\varrho}$ denotes time derivative, $\gamma$ is the
damping factor and the Lindblad superoperator acts as follows
$\mathcal L[a]\varrho= 2a^\dagger \varrho a -a^\dagger
a\varrho-\varrho a^\dagger a$.  Upon writing $\eta = e^{- \gamma
t}$ the solution of the Master equation can be written as
\begin{align}
\varrho(\eta) & = \sum_m V_m\:\varrho\: V_m^\dag \\
V_m & = [(1-\eta)^m/m!]^{\frac12} a^m\:\eta^{\frac{1}{2}(a^{\dag}a-m)}\,,
\nonumber
\end{align}
where $\varrho$ is the initial state.
In particular for the system initially prepared in a Fock state
$\varrho_p=|p\rangle\langle p|$, we obtain, after evolution, the
mixed state
\begin{align}
\varrho_p(\eta) &= \sum_m V_m \varrho_p V_m^{\dag} =
\sum_{l=0}^{p} \alpha_{l,p}(\eta) |l\rangle\langle l| \label{eq:loss}
\end{align}
with  $\alpha_{l,p}(\eta) = \binom{p}{l} (1-\eta)^{p-l} \eta^l$.
The reference Gaussian state corresponding to $\varrho_p (\eta)$
is a thermal state $\tau_p(\eta) = \nu (p\eta)$ with average
photon number $p\eta$.  Non-Gaussianity of $\varrho_p(\eta)$
can be evaluated analytically, we have
\begin{align}
\delta_{p\eta} &\equiv \delta [\varrho_p (\eta)] \nonumber \\
&= \frac{1}{2(1 - \eta)^{2m}\: {}_2 F_1\left(-m,-m,1;\frac{\eta^2}
{(\eta -1)^2}\right)} \nonumber \\
&\times \left\{(1 - \eta)^{2m}\: {}_2 F_1\left(-m,-m,1;\frac{\eta^2}
{(\eta -1)^2}\right) \right. \nonumber \\
&\left.  + \: (1 + 2m\eta)^{-1} - \frac{2(1 + (m-1)\eta)^m}
{(1+m\eta)^{m+1}} \right\}
\end{align}
${}_2 F_1(a,b,c;x)$ being a hypergeometric function.
We show the behavior of $\delta_{p\eta}$ in
Fig. \ref{f:procs} as a function of $1-\eta$ for different
values of $p$.  As it is apparent from the plot $\delta_{p\eta}$
is a monotonically decreasing function of $1- \eta$ as well as a
monotonically increasing function of $p$. That is, at fixed time
$t$ the higher is the initial photon number $p$, the larger is
the resulting non-Gaussianity.
\par
Let us now consider the de-Gaussification protocol obtained by
the process of photon subtraction. Inconclusive Photon
Subtraction (IPS) has been introduced for single-mode and
two-mode states in \cite{IPS1,IPS2a,IPS2b} and experimentally
realized in \cite{IPS_Wenger}. In the IPS protocol an input state
$\varrho^{(in)}$ is mixed with the vacuum at a beam splitter (BS)
with transmissivity $T$ and then, on/off photodetection with
quantum efficiency $\epsilon$ is performed on the reflected beam.
The process can be thus characterized by two parameters: the
transmissivity $T$ and the detector efficiency $\epsilon$. Since
the detector can only discriminate the presence from the absence
of light, this measurement is inconclusive, namely it does not
resolve the number of detected photons. When the detector clicks,
an unknown number of photons is subtracted from the initial state
and we obtain the conditional IPS state $\varrho_{IPS}$. The
conditional map induced by the measurement is
non-Gaussian \cite{IPS2b}, and the output state is
de-Gaussified. Upon applying the IPS protocol to the (Gaussian)
single-mode squeezed vacuum $S(r)|0\rangle$ ($r \in
\mathbbm{R}$), where $S(r)$ is the real squeezing operation we
obtain \cite{IPS1} the conditional state $\varrho_{IPS}$, whose
characteristic function $\chi[\varrho_{IPS}](\lambda)$ is a sum
of two Gaussian functions and therefore is no longer Gaussian.
The corresponding Gaussian reference state is a squeezed thermal
state $\tau_{IPS}=S(\xi_{IPS})\nu (N_{IPS}) S^{\dag}(\xi_{IPS})$
where the parameters $\xi_{IPS}$ and $N_{IPS}$ are analytic
functions of $r$, $T$ and $\epsilon$. Non-Gaussianity
$\delta_{IPS}=\delta_{IPS}(T,\epsilon,r)$ has been evaluated, and
in Fig. \ref{f:procs} (right) we report $\delta_{IPS}$ for $r=0.5$ as a
function of the transmittivity $T$ for different values of the
quantum efficiency $\epsilon$. As it is apparent from the plot
the IPS protocol indeed de-Gaussifies the input state, {\em i.e.}
nonzero values of the non-Gaussianity are obtained. We found that
$\delta_{IPS}$ is an increasing function of the transmissivity
$T$ which is the relevant parameter, while the quantum efficiency
$\epsilon$ only slightly affects the non-Gaussian character of
the output state. The highest value of non-Gaussianity is
achieved in the limit of unit transmissivity and unit quantum
efficiency
$$\lim_{T,\eta\rightarrow 1}\delta_{IPS}=\delta[|1\rangle\langle 1|]
= \delta[S(r)|1\rangle\langle 1|S^{\dag}(r)],
$$
where the last equality is derived from Lemma 2. This result is
in agreement with the fact that a squeezed vacuum state
undergoing the IPS protocol is driven towards the target state
$S(r)|1\rangle$ in the limit of $T,\epsilon\rightarrow 1$
\cite{IPS1}. Finally, we notice that for $T,\epsilon \neq 1$ and
for $r\rightarrow\infty$ the non-Gaussianity vanishes.  In turn,
this corresponds to the fact that one of the coefficients of the
two Gaussians of $\chi[\varrho_{IPS}](\lambda)$ vanishes, {\em
i.e.} the output state is again a Gaussian one.
\section{Conclusion and outlooks}
\label{s:out}
Having at disposal a good measure of non-Gaussianity for quantum
state allows us to define a measure of the non-Gaussian character
of a quantum operation. Let us denote by ${\cal G}$ the whole
set of Gaussian states. A convenient definition for the
non-Gaussianity of a map ${\cal E}$ reads as follows
$\delta[{\cal E}] = \max_{\varrho \in {\cal G}} \delta [{\cal E}
(\varrho)]$, where ${\cal E} (\varrho)$ denotes the quantum state
obtained after the evolution imposed by the map. Indeed, for
a Gaussian map ${\cal E}_g$, which transforms any input Gaussian
state into a
Gaussian state, we have $\delta[{\cal E}_g]=0$. Work along this
line is in progress and results will be reported elsewhere.
\par
In conclusion, we have proposed a measure of the non-Gaussian
character of a CV quantum state. We have shown that our measure
satisfies the natural properties expected from a good measure of
non-Gaussianity, and have evaluated the non-Gaussianity of some
relevant states, in particular of states undergoing
Gaussification and de-Gaussification protocols. Using our measure
an analogue non-Gaussianity measure for quantum operations may be
introduced.
\section*{Acknowledgments}
This work has been supported by MIUR project PRIN2005024254-002,
the EC Integrated Project QAP (Contract No.\ 015848) and Polish
MNiSW grant 1~P03B~011~29.
\appendix
\section{Gaussian reference with unconstrained mean value}
As we have seen from the above examples $\delta[\varrho]$ of
Eq. (\ref{eq:nonG}) represents a good measure of the non-Gaussian
character of a quantum state. A question arises on whether
different choices for the reference Gaussian state $\tau$ may
lead to alternative, valid, definitions. As for example (for
single-mode states) we may define
\begin{align}
\delta^{\prime}[\varrho] &=
\min_{\tau} \D^2_{HS}[\varrho,\tau]/ \mu[\varrho], \label{eq:nonG2}
\end{align}
where $\tau=D(C)S(\xi)\nu (N) S^{\dag}(\xi)D^{\dag}(C)$ is a Gaussian state
with the same covariance matrix of $\varrho$ and unconstrained
vector of mean values $\X=( \hbox{Re} C, \hbox{Im} C)$ used to minimize the
Hilbert-Schmidt distance. Here we report few examples of the comparison
between the results already obtained using (\ref{eq:nonG}) with that
coming from (\ref{eq:nonG2}).
As we will see either the two definitions coincide or $\delta'$ and $\delta$
are monotone functions of each other.  Since the
definition (\ref{eq:nonG}) corresponds to an easily computable
measure we conclude that it represents the most convenient
choice.
\par
Let us first consider the Fock state $\varrho=|p\rangle\langle p|$. According to
(\ref{eq:nonG2}), the reference Gaussian state is given by a displaced thermal
states $\tau^{\prime}= D(C)\nu_pD^{\dag}(C)$. The overlap between $\varrho$
and $\tau^\prime$ is given by
\begin{align}
\kappa[|p\rangle\langle p|,\tau^\prime] & =
\frac{1}{1+p}\exp\left\{-\frac{C^2}{1+p}\right\}
\left(\frac{p}{1+p}\right)^p \nonumber \\
\:\:\:&\times L_p\left(-\frac{C^2}{p(1+p)}\right) \label{eq:OverFock2}
\end{align}
The maximum of (\ref{eq:OverFock2}) is achieved for $C=0$, which coincides
with the assumptions $C=\Tr[a|p\rangle\langle p|]$.
\par
Let us consider the quantum state (\ref{eq:loss})
obtained as the solution of the loss Master Equation for an initial
Fock state $|p\rangle\langle p|$. The unconstrained Gaussian reference is
again a displaced thermal state $\tau^\prime =D(C)\nu_{p\eta}D^{\dag}(C)$,
and the overlap is given by
\begin{align}
\kappa[\varrho_p(\eta),\tau^\prime] &= \Tr[\tau \varrho_p(\eta)]
= \frac{\left( 1+ \eta (p-1)
\right)^p}{(1+ p\eta)^{p+1}} \nonumber \\ & \times L_p\left(
\frac{\eta |C|^2}{(1+p\eta)(\eta(1-p) - 1)}
\right)\, e^{- \frac{|C|^2}{1+p\eta}}
\nonumber
\end{align}
Again, since the overlap is maximum for $C=\Tr[a\varrho_p(\eta)]=0$, both
definitions give the same results for the non-Gaussianity.
\par
\begin{figure}
\centering
\begin{tabular}{c}
\includegraphics[width=0.4\textwidth]{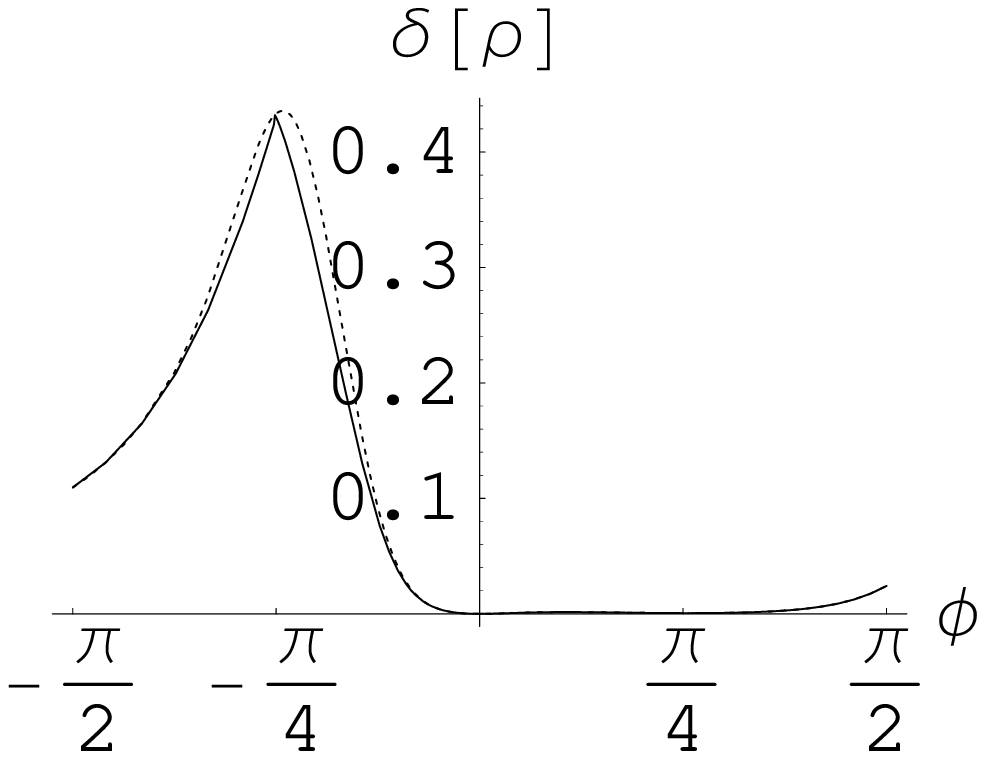}\\
\includegraphics[width=0.4\textwidth]{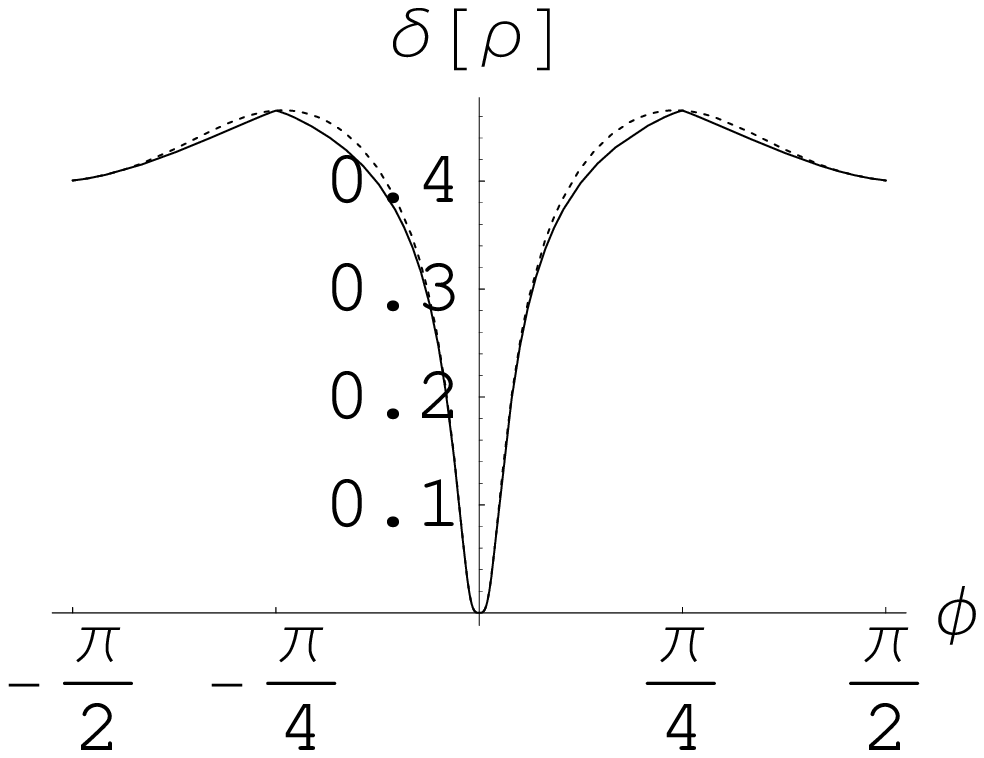}
\end{tabular}
\caption{Non-Gaussianity of a Schr\"odinger cat-like state as a
function of the superposition parameter $\phi$, with either $C$
obtained by numerical minimization (solid) or with $C=\hbox{Tr}[a\varrho]$
(dotted). (Left): $\alpha=0.5$; (Right): $\alpha= 5$.} \label{fig:f4}
\end{figure}

Let us now consider the Schr\"odinger cat-like states of (\ref{eq:cat}).
The reference Gaussian state is a displaced squeezed thermal state, with
squeezing and thermal photons as calculated before.
The optimization over the free parameter $C$ may be done numerically.
In Fig. \ref{fig:f4} we show the
non-Gaussianitiy, both as resulting from (\ref{eq:nonG2}) and by choosing
$C=\Tr[a\varrho_S]$ as in (\ref{eq:nonG}), as
a function of $\epsilon$. The two curves are almost the same, with
no qualitative differences.

\end{document}